\documentclass[aps,prd,showpacs,nofootinbib]{revtex4}
\usepackage{bbm}
\usepackage{mathrsfs}
\usepackage{epsfig,psfrag}
\usepackage{amsmath,amsfonts,amssymb}
\usepackage[usenames]{color}
\usepackage{bm}
\usepackage{ulem}
\begin{document}

\title{Schwarzschild Quantum Fluctuations from Regge-Wheeler Scattering}
 
\author{Noah Graham}
\email{ngraham@middlebury.edu}
\affiliation{Department of Physics,
Middlebury College,
Middlebury, VT 05753  USA}

\pacs{03.65.Nk,11.80.-m,04.62.+v}

\begin{abstract}
We apply a multichannel variable phase method to scattering from
Regge-Wheeler potentials.  Using a reduced version of the WKB
subtraction developed by Candelas and Howard, this approach allows for
efficient numerical calculations of scattering data for imaginary wave
number, making it possible to compute quantum expectation values in a
Schwarzschild curved spacetime background through Wick rotation to the
imaginary frequency axis.  These scattering theory techniques are also
potentially applicable to a variety of other problems involving wave
propagation in curved spacetime.
\end{abstract}

\maketitle

\section{Introduction}

The Regge-Wheeler formalism \cite{PhysRev.108.1063} expresses
fluctuations propagating in a Schwarzschild spacetime
background in terms of ordinary scattering from a family of asymmetric
one-dimensional potentials, one for each partial wave.  This approach
can be used to analyze many aspects of wave propagation in the
Schwarzschild geometry, including  orbits, wave scattering, and
quasi-normal modes (see for example
Refs.\ \cite{Sanchez:1976fcl,Sanchez:1976xm,Sanchez:1977sia,
PhysRevLett.52.1361,PhysRevD.31.1869,
PhysRevD.46.4477,PhysRevD.52.1808,lrr-1999-2,
Motl:2003cd,doi:10.1063/1.1626805,PhysRevD.74.064005,
PhysRevD.84.104002,Konoplya:2011qq,PhysRevD.86.104060,Raffaelli2013}).
By summing over fluctuation spectra, one can use these results
to obtain quantum expectation values.  For example, the luminosity of
Hawking radiation can be expressed in terms of the transmission
coefficient of Regge-Wheeler scattering \cite{Candelas}.  Calculations
of local densities can then allow for detailed exploration of the
black hole ``atmosphere,'' the region extending significantly beyond
the event horizon from which Hawking radiation originates (see for
example Ref.\ \cite{Giddings:2015uzr}).

While it is fairly straightforward to obtain expressions for
quantum expectation values of local densities in terms of the
fluctuation Green's function, in curved spacetime backgrounds these
calculations typically become finite only as a result of numerically
ill-conditioned cancellations.  As a result, it is common to develop
approximate expansions that can either be subtracted and then added
back in analytically to improve the convergence of the calculation, or
simply used on their own \cite{Candelas,PhysRevD.25.1499,Candelas2,
PhysRevLett.53.403,PhysRevD.39.1130,Anderson,PhysRevLett.70.1739,
PhysRevD.51.4337,PhysRevD.62.064007,PhysRevD.67.044021,PhysRevD.91.104028}.
Here we take an approach that makes use of simple forms of these
expansions, particularly those developed in Ref.\ \cite{Candelas2},
combined with computational techniques to compute scattering data for
Regge-Wheeler potentials.  The result provides a more generic
framework for calculations of this kind, at the cost of
greater demands on the numerical computation.  We demonstrate this
approach in typical quantum field theory calculations, focusing on the
expectation values $\langle \Phi^2 \rangle_B$ and
$\langle \dot \Phi^2 \rangle_B$ for a massless scalar field in the
Boulware vacuum of Schwarzschild spacetime.

These techniques make it possible to compute scattering data for
complex wave number, which we take advantage of in Wick rotating the
quantum field theory calculation to the imaginary frequency axis
$k=i\kappa$.  This capability is also potentially applicable to
problems formulated directly on the imaginary axis, such as
calculations of quasi-normal modes and problems in thermal field
theory at temperature $T$, where the integral over wave number $k$ is
replaced by a sum over Matsubara modes given by $k = 2\pi i n T$ for
integer $n$.  The latter case tends to arise naturally in analyzing
vacua in black hole spacetimes with different thermal boundary conditions.

\section{Scattering From Regge-Wheeler Potentials}

We consider a free massless scalar field $\Phi$ propagating in a
Schwarzschild metric background, so that its dynamics are given by the
Klein-Gordon equation in curved spacetime
\begin{equation}
g^{\mu \nu} \nabla_\mu \nabla_\nu \Phi = 0 \,,
\end{equation}
where $g_{\mu \nu} = {\rm diag}(-f(r), f(r)^{-1}, r^2, r^2 \sin^2 \theta)$
with
\begin{equation}
f(r) = 1 - \frac {2 GM}{r}
\end{equation}
is the Schwarzschild metric in the coordinates $t$, $r$, $\theta$, and
$\varphi$, and $\nabla_\mu$ is the associated covariant derivative.

Because the system is symmetric under both rotation and time
translation, we can separate variables in the Klein-Gordon equation to
obtain solutions of the form
\begin{equation}
\Phi_\ell(k, r) = \frac{1}{\sqrt{4 \pi k}} \psi_\ell(k,r) e^{-ik t}
Y_\ell^m(\theta, \varphi)\,,
\end{equation}
where we have introduced the usual spherical harmonic
functions $Y_\ell^m(\theta, \varphi)$ and chosen convenient
normalization conventions.  These solutions are then given in terms of
the angular quantum numbers $\ell$ and $m$ and the frequency
$k=\omega$.  (We work in natural units where the speed of light 
$c$ and the reduced Planck constant $\hbar$ are both equal to one, but
include Newton's constant $G$ explicitly.)  Analytic continuation in
$k$ will play a key role in the subsequent analysis.

\begin{figure}[htbp]
\includegraphics[width=0.45\linewidth]{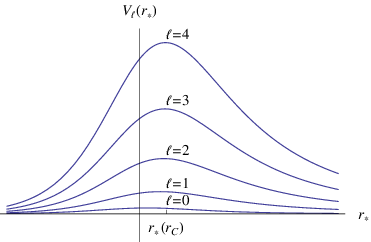}
\caption{Regge-Wheeler potentials $V_\ell(r_*)$.}
\label{fig:potentials}
\end{figure}

After some algebra making use of the standard Schwarzschild metric 
and connection, we obtain the radial equation
\begin{equation}
-\frac{f(r)}{r^2} \frac{\partial}{\partial r} \left(r^2
f(r) \frac{\partial \psi_\ell}{\partial r} \right)
+ \frac{f(r) \ell (\ell+1)}{r^2} \psi_\ell(k,r) = k^2 \psi_\ell(k,r)
\label{eqn:KG}
\end{equation}
for fluctuations of the $\Phi$ field.  We restrict our attention to
the region outside the event horizon, $r>2 GM$, and introduce the
Regge-Wheeler ``tortoise'' radial coordinate $r_*$, given by
\begin{equation}
r_* = r + 2 GM \log \left(\frac{r}{2GM} -1 \right)
\qquad r = 2GM \left(1 + {\rm W}\left(e^{r_*/(2GM) - 1}\right)\right) \,,
\end{equation}
where ${\rm W}(z)$ is the Lambert product logarithm function
\cite{lambert1758observationes,euler1783serie,corless1996lambertw}.
Note that $r_*$ is defined on the whole real line, with $r_* \to
-\infty$ corresponding to $r\to 2GM$ and $r_* \to +\infty$
corresponding to $r \approx r_*$.  It obeys
\begin{equation}
\frac{dr}{dr_*} = f(r) \,,
\end{equation}
so that the Klein-Gordon equation becomes
\begin{equation}
-\frac{\partial^2 \phi_\ell}{\partial r_*^2} + 
V_\ell(r_*) \phi_\ell(k,r_*) = k^2 \phi_\ell(k,r_*) \,,
\label{eqn:KG2}
\end{equation}
where $\psi_\ell(k,r_*) = \phi_\ell(k,r_*)/r$, and the Regge-Wheeler
potential is given by
\begin{equation}
V_\ell(r_*) = \left(1-\frac{2 G M}{r}\right)\left(\frac{\ell(\ell+1)}{r^2}
+ \frac{2 G M}{r^3}\right) \,.
\end{equation}
In these expressions $r$ is now a function of $r_*$, as
given above.  For each $\ell$, we have thus obtained a set of ordinary
one-dimensional scattering problems in the repulsive ``centrifugal''
potential $V_\ell(r_*)$.  The first few such potentials are shown in
Fig.\ \ref{fig:potentials}.  As $\ell$ increases, this potential
grows in height, and the location of its maximum approaches the radius
of the circular photon orbit at $r= r_C = 3 GM$.

One may immediately be concerned that one-dimensional scattering on
the real line represents a two-channel problem, while an individual
partial wave in three dimensions represents only a single scattering
channel.  This discrepancy is addressed by imposing appropriate
boundary conditions at $r_*=\pm \infty$.  In the study of black hole
thermodynamics, it is instructive to compare results for vacuum states
defined through different choices of these boundary conditions;
particular cases of interest are the Boulware, Unruh, and
Hartle-Hawking vacua.  The Boulware vacuum corresponds to a
state with no radiation at infinity, but with divergent energy
density at $r=2GM$, while the Unruh and Hartle-Hawking vacua avoid this
pathology at the expense of introducing radiation at $r=\infty$.  The
Unruh vacuum corresponds to our expectation for a physical black hole,
with an outgoing flux of Hawking radiation at infinity, while the
Hartle-Hawking vacuum corresponds to a black hole with outgoing
radiation at $r=\infty$ in equilibrium with a corresponding external
source of incoming radiation at the same temperature.  As a result, we
will consider the full one-dimensional scattering problem and later
project onto the relevant subspace for a particular problem.

In the standard approach (see for example 
Refs.\ \cite{Candelas,Candelas2,Anderson,PhysRevD.91.104028}), one
defines the usual reflection and transmission components of
one-dimensional scattering for two channels representing incoming
waves from the left and right with the asymptotic behavior
\begin{equation}
\overrightarrow{\phi_\ell}(k, r_*) = 
\left\{ \begin{array}{l@{\quad}l}
e^{i k r_*} + \overrightarrow{A_\ell} e^{-i k r_*} & r_* \to -\infty \cr
B_\ell e^{i k r_*}   & r_* \to +\infty
\end{array} \right.
\qquad \hbox{and} \qquad
\overleftarrow{\phi_\ell}(k, r_*) = 
\left\{ \begin{array}{l@{\quad}l}
B_\ell e^{-i k r_*}   & r_* \to -\infty \cr
e^{-i k r_*} + \overleftarrow{A_\ell} e^{i k r_*} & r_* \to +\infty 
\end{array} \right. \,,
\label{eqn:regs}
\end{equation}
where the transmission coefficient $B_\ell$ is the same in both cases.
(There has also been more recent work \cite{Gray:2015xig} applying a
formalism based on path-ordered exponentials and transfer matrices.)

For numerical calculations, we will apply methods of analytic
scattering theory based on the variable phase approach
\cite{variable,Graham:2002xq}.  We will first show how these techniques
apply to  scattering in the standard basis, and then discuss
modifications of this approach that are advantageous for our problem.
To begin, we choose an arbitrary ``origin'' $z_{*,\ell}$.  We then define
the outgoing wave solution $f_{+,\ell}(k,r_*)$, which approaches 
$e^{i k r_*}$ as $r_* \to +\infty$, and the outgoing wave solution
$f_{-,\ell}(k,r_*)$, which approaches $e^{-i k r_*}$ as $r_* \to -\infty$.
To apply the variable phase method, we parameterize $f_{\pm,\ell}(k,r_*) =
g_{\pm,\ell} (k,r_*) e^{\pm i k r_*}$ and solve for $g_{\pm,\ell} (k,r_*)$ by
integrating Eq.\ (\ref{eqn:KG2}) inward from $r_* = \pm \infty$, with
the boundary condition that  $g_{\pm,\ell} (k,r_*)$ goes to the identity at
$r_*=\pm \infty$.  Since the functions
$f_{\pm,\ell}(-k,r_*)=f_{\pm,\ell}(k,r_*)^*$
also solve the same differential equation, we have a total of four
incoming and outgoing wave solutions.  To form regular solutions as in
Eq.\ (\ref{eqn:regs}), we impose continuity of the wavefunction and
first derivative as boundary conditions at $r_* = z_{*,\ell}$, together with
one additional boundary condition at infinity in each channel:  for the
regular solution representing scattering of a wave incoming from the
left, we require that there be no incident wave from the right, and,
similarly, for the regular solution representing scattering of a wave
incoming from the right, we require that there be no incident wave from
the left.

The equation for $g_{\pm,\ell}(k,r_*)$ becomes
\begin{equation}
- g_{\pm,\ell}''(k,r_*) - 2ik g_{\pm,\ell}(k,r_*) + v_\ell(r_*)
  g_{\pm,\ell}(k,r_*) = 0 \,,
\end{equation}
with boundary conditions at large $|R_*|$,
\begin{eqnarray}
g_{+,\ell}(k,R_*) = \frac{i^{\ell+1}k R_* h^{(1)}_\ell(k R_*)}{ e^{ik R_*}} 
&& g_{+,\ell}'(k,R_*) = \frac{i^{\ell+1}k \left(
(1 - i k R_* + \ell) h^{(1)}_\ell(k R_*) - k R_* h^{(1)}_{\ell+1}(k R_*)
\right)}{e^{ik R_*}}
\cr
g_{-,\ell}(k,-R_*) = 1
&& g_{-,\ell}'(k,-R_*) = 0 \,,
\end{eqnarray}
where the values for $g_{+,\ell}(k,R_*)$ represent an improved estimate
taking into account the leading behavior of the potential at large
$r$, as described in the Appendix.  To improve the numerical precision
of the calculation, it is also helpful to take advantage of the
Wronskian relations between the solutions in each region,
\begin{equation}
g_{\pm,\ell}(k,z_{*,\ell}) g_{\pm,\ell}'(-k,z_{*,\ell})-
g_{\pm,\ell}'(k,z_{*,\ell}) g_{\pm,\ell}(-k,z_{*,\ell}) = 
2 i k \left(g_{\pm,\ell}(k,z_{*,\ell})
g_{\pm,\ell}(-k,z_{*,\ell})-1\right) \,,
\end{equation}
evaluated at the origin $z_{*,\ell}$.

\begin{figure}
\includegraphics[width=0.45\linewidth]{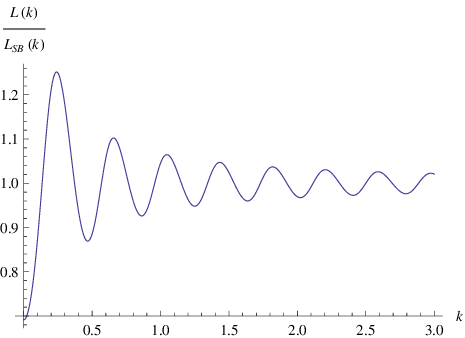}
\caption{For a massless real scalar field, luminosity of Hawking
radiation $L(k)$ at frequency $k$ compared to the luminosity of
Stefan-Boltzmann radiation  $L_{\rm SB}(k)$ from a blackbody sphere of
radius $r = 3\sqrt{3} GM$ and temperature $T = \frac{1}{8\pi GM}$.
The ratio is plotted as a function of $k$, in units where $2 GM = 1$.}
\label{fig:luminosity}
\end{figure}

Using this formalism, we can compute the outgoing wave
solutions $f_{\pm,\ell}(k,r_*)$ and the regular solutions
$\overleftarrow{\phi_\ell}(k, r_*)$ and
$\overrightarrow{\phi_\ell}(k, r_*)$, along with the $S$-matrix
\begin{equation}
\hat S_\ell(k) = \begin{pmatrix} B_\ell & \overrightarrow{A_\ell} \cr 
\overleftarrow{A_\ell} & B_\ell \end{pmatrix}\,,
\end{equation}
which is unitary for real $k$, and the corresponding eigenphase
shifts, which are the eigenvalues of $\frac{1}{2i} \log
\hat S$.  For thermal quantities, the exponential damping due to the
Boltzmann factor tends to eliminate subtleties involved with
potentially divergent quantities.  For example, we can obtain the
luminosity of Hawking radiation at infinity as \cite{Candelas}
\begin{equation}
L_{\rm tot} = \int_0^\infty L(k) \, dk
= \frac{1}{2\pi} \int_0^\infty k
\sum_{\ell=0}^\infty \frac{(2\ell+1)|B_\ell(k)|^2}{\exp(k/T) - 1} dk \,,
\label{eqn:luminosity}
\end{equation}
where $T=\frac{1}{8\pi GM}$ is the Hawking temperature.
As shown in Fig.\ \ref{fig:luminosity}, the variable phase approach
yields a calculation that agrees with standard results; roughly, it
describes a spectrum that corresponds to the luminosity of
Stefan-Boltzmann radiation
\begin{equation}
L_{\rm SB, tot} = \int_0^\infty L_{\rm SB}(k) \, dk
= \int_0^\infty \frac{27 (GM)^2 k^3\, dk}
{2\pi(e^{8\pi GM k}-1)} = \frac{9}{40960\pi (GM)^2}\,,
\end{equation}
at temperature $T$ emitted by a sphere of radius $r = 3\sqrt{3} GM$, the
geometric-optics limit for the scattering cross section, that is then
modified by ``gray-body'' factors representing the scattering of the
outgoing radiation in the Schwarzschild background.

\section{Quantum Expectation Values in Curved Spacetime}

Renormalized densities can be expressed in terms of sums over
the corresponding mode functions.  For example, Ref.\ \cite{Candelas}
computes the expectation value of $\Phi^2$ in the Boulware vacuum as 
\begin{equation}
\langle\Phi(r)^2\rangle_B = \frac{1}{16\pi^2 r^2} \int_0^\infty
\frac{dk}{k}
\left[\sum_{\ell=0}^\infty (2\ell + 1)\left(
|\overrightarrow{\phi}_\ell(k,r)|^2 + |\overleftarrow{\phi}_\ell(k,r)|^2
\right) - \frac{4 k^2 r^2}{1-\frac{2 G M}{r}}\right]
-\frac{(GM)^2}{48 \pi^2 r^4 (1-\frac{2 G M}{r})}
\label{eqn:phi2}
\end{equation}
where the last two terms represent renormalization counterterms, which are
found by point-splitting \cite{PhysRevD.14.2490}.  The last term is a
finite renormalization that arises from the expansion
\begin{equation}
\frac{1}{8 \pi^2 \sigma(\epsilon)}  = 
\frac{1}{4\pi^2}\left[
\frac{1}{\left(1-\frac{2GM}{r}\right)} \frac{1}{\epsilon^2}
+ \frac{(GM)^2}{12 r^4\left(1-\frac{2GM}{r}\right)}
+ \frac{(GM)^3(4r - 7GM)}
{240 r^8\left(1-\frac{2GM}{r}\right)} \epsilon^2
+ {\cal O} (\epsilon^4)\right] \,,
\label{eqn:interval}
\end{equation}
where $\sigma(\epsilon)$ is the invariant distance squared between the
infinitesimally separated points $(t,r,\theta,\varphi)$ and
$(t+\epsilon,r,\theta,\varphi)$.  The ${\cal O}(\epsilon^2)$ term
in Eq.\ (\ref{eqn:interval}) is not needed for the present
calculation, but is included for use below.  This expansion is obtained
by writing $\sigma(\epsilon) = \frac{1}{2}\sigma_\mu(\epsilon) 
\sigma^\mu(\epsilon)$, with \cite{PhysRevD.51.4337}
\begin{equation}
\sigma^\mu(\epsilon) = 
\delta_t^\mu \epsilon
- \frac{1}{2}\Gamma_{tt}^\mu \epsilon^2
+ \frac{1}{6} \Gamma_{tt}^\rho \Gamma_{\rho t}^\mu \epsilon^3
- \frac{1}{24} \Gamma_{tt}^\rho \left(
R^\mu_{t\rho t} + \Gamma_{\rho t}^\lambda \Gamma_{\lambda t}^\mu
\right)\epsilon^4
+ \frac{1}{120} \Gamma_{tt}^r \left(
2 \Gamma_{rt}^\mu R^r_{trt} 
- \Gamma^r_{tt} R^\mu_{rtr}
+ \Gamma^r_{tt} \Gamma_{rt}^t \Gamma_{rt}^\mu
\right) \epsilon^5 + {\cal O} (\epsilon^6) \,, 
\end{equation}
where $\Gamma_{\alpha \beta}^\gamma$ and 
$R^\lambda_{\alpha \beta \gamma}$ represent the components of the
Schwarzschild connection and curvature respectively.  The point
splitting method then gives the result in Eq.\ (\ref{eqn:phi2}) as the
limit \cite{Candelas}
\begin{equation}
\langle\Phi(r)^2\rangle_B = 
\lim_{\epsilon\to 0} \left[
\int_0^\infty \frac{dk}{k}
e^{-ik\epsilon} \frac{1}{16 \pi^2 r^2}
\sum_{\ell=0}^\infty (2\ell + 1)\left(
|\overrightarrow{\phi}_\ell(k,r)|^2 + |\overleftarrow{\phi}_\ell(k,r)|^2
\right) - \frac{1}{8 \pi^2 \sigma(\epsilon)}\right] \,.
\label{eqn:pointsplit}
\end{equation}

As in Ref.\ \cite{Graham:2002xq}, the squared wavefunctions can be
re-expressed in terms of the Green's function at coincident points, 
\begin{equation}
|\overrightarrow{\phi}_\ell(k,r)|^2 + |\overleftarrow{\phi}_\ell(k,r)|^2
= 2k {\, \rm Im\,} G_\ell(r,r,k) \,,
\end{equation}
which in turn can be obtained as a product of the regular and outgoing
solutions normalized by the Jost function, yielding a form that is
amenable to analytic continuation into the upper half-plane.

For numerical calculations of quantities like those in 
Eq.\ (\ref{eqn:phi2}), however, we can anticipate
difficulties that arise as a combination of the usual field theory
divergences together with the growing strength of the Regge-Wheeler
potential for increasing $\ell$.  In particular, the sum over partial
waves of the Green's function yields a quadratic divergence in $k$,
which then cancels with the first renormalization counterterm.  The
remaining integral is in principle convergent, but in practice
converges only as a result of the cancellation of oscillating
contributions in $k$.  It is thus not amenable to numerical
calculation, but rather can be analyzed in terms of asymptotic limits
\cite{Candelas} in which one can extract leading analytic behavior.

Motivated by the Born subtraction approach in ordinary quantum field
theory \cite{Graham09}, we aim to improve this situation by implementing
local subtractions that then allow for a Wick rotation to imaginary
wave number; as indicated above, the behavior of the Green's function
in the complex $k$-plane is of interest for its own sake as well. 
We begin by defining the local spectral density
\begin{equation}
\rho_\ell(k,r) = \frac{2k}{i} G_\ell(r,r,k) \,,
\label{eqn:specden}
\end{equation}
so that for $k$ real, 
\begin{equation}
{\, \rm Re \,} \rho_\ell(k,r) = 
|\overrightarrow{\phi}_\ell(k,r)|^2 + |\overleftarrow{\phi}_\ell(k,r)|^2 \,.
\end{equation}
While these definitions are written in terms of $r$, in practice we
will compute the scattering data in terms of $r_*$ and then convert to
the corresponding $r$.
Using this definition, we rewrite Eq.\ (\ref{eqn:phi2}) as
\begin{eqnarray}
\langle\Phi(r)^2\rangle_B &=& \frac{1}{16\pi^2 r^2} \int_0^\infty
\frac{dk}{k} \left[\sum_{\ell=0}^\infty (2\ell + 1) {\, \rm Re \,}
\rho_\ell(k,r)
 - \frac{4 k^2 r^2}{1-\frac{2 G M}{r}}\right]
-\frac{(GM)^2}{48 \pi^2 r^4 (1-\frac{2 G M}{r})} \cr
&=& \frac{1}{32\pi^2 r^2} \int_{-\infty}^\infty
\frac{dk}{\sqrt{k^2}} \left[\sum_{\ell=0}^\infty (2\ell + 1) \rho_\ell(k,r)
- \frac{4 k^2 r^2}{1-\frac{2 G M}{r}}\right]
-\frac{(GM)^2}{48 \pi^2 r^4 (1-\frac{2 G M}{r})} \,,
\label{eqn:kaxis}
\end{eqnarray}
where we have used that $\rho_\ell(-k,r_*) =
\rho_\ell(k,r_*)^*$.  However, we must be careful to note a subtlety
of this calculation:  while the sum over $\ell$ of the real part of
$\rho_\ell(k, r_*)$ converges, the sum over the imaginary part does
not, even though the contribution from
each term cancels between $k$ and $-k$.  If we
consider the large-$\ell$ limit by setting $k=0$ in
Eq.~(\ref{eqn:KG}), we find 
\cite{Candelas,Candelas2,Anderson,PhysRevD.91.104028}
that this equation has solutions given by the Legendre functions
$P_\ell(\xi)$ and $Q_\ell(\xi)$, with $\xi = \frac{r}{GM} - 1$.  Using
\begin{equation}
P_\ell(z) Q_\ell(z) \approx \frac{1}{\sqrt{z^2-1} (2\ell+1)}
\hbox{\qquad for $\ell$ large with $z>1$ real,}
\end{equation}
and normalizing the solutions using the Wronskian
\begin{equation}
P_\ell(z) Q_\ell'(z) - P_\ell'(z) Q_\ell(z) = \frac{1}{1-z^2} \,,
\end{equation}
we find that the summand in Eq.\ (\ref{eqn:kaxis}) approaches the
imaginary quantity $\frac{4kr}{i\sqrt{1-\frac{2 GM}{r}}}$
for large $\ell$.  Thus we can rewrite Eq.\ (\ref{eqn:kaxis}) as
\cite{Candelas,Candelas2,Anderson,PhysRevD.91.104028}
\begin{eqnarray}
\langle\Phi(r)^2\rangle_B
&=& \frac{1}{32\pi^2 r^2} \int_{-\infty}^\infty
\frac{dk}{\sqrt{k^2}} \left[\sum_{\ell=0}^\infty 
(2\ell + 1) \left(\rho_\ell(k,r) - \frac{4k r^2}{i GM}
P_\ell(\xi) Q_\ell(\xi)\right)
- \frac{4 k^2 r^2}{1-\frac{2 G M}{r}}\right]
-\frac{(GM)^2}{48 \pi^2 r^4 (1-\frac{2 G M}{r})}  \cr
&=& \frac{1}{32\pi^2 r^2} \int_{-\infty}^\infty
\frac{dk}{\sqrt{k^2}} \left[\sum_{\ell=0}^\infty \left(
(2\ell + 1) \rho_\ell(k,r) - \frac{4kr}{i\sqrt{1-\frac{2 GM}{r}}}
\right)
- \frac{4 k^2 r^2}{1-\frac{2 G M}{r}}\right]
-\frac{(GM)^2}{48 \pi^2 r^4 (1-\frac{2 G M}{r})} \,,
\label{eqn:kaxis2}
\end{eqnarray}
where we have used the identity \cite{Candelas2}
\begin{equation}
\sum_{\ell=0}^\infty \left[(2\ell+1) P_\ell(\xi) Q_\ell(\xi) -
\frac{1}{\sqrt{\xi^2-1}}\right] = 0
\end{equation}
to equate these two expressions.  The subtraction we have introduced 
in Eq.\ (\ref{eqn:kaxis2}) does
not change our original result, since it is purely imaginary and its
contribution cancels between $-k$ and $+k$, but the
the sums over $\ell$ in Eq.\ (\ref{eqn:kaxis2}) are now convergent.

The subtraction linear in $k$ in Eq.~(\ref{eqn:kaxis2})
cancels the quadratic divergence in the mode sum.  However, 
this cancellation of large quantities is still ill-behaved
numerically; we would strongly prefer to implement this subtraction
term by term.  We do so by adding and subtracting the leading WKB
approximation \cite{Candelas2,MaassenvandenBrink:2003as}\footnote{We
can identify the Matsubara mode $n$ used in Ref.\ \cite{Candelas2}
with the frequency $k$ used here via $k = \frac{in}{4 GM}$.}
\begin{equation}
\rho_\ell^{\rm WKB}(k,r) = \frac{2r^2 k}{i GM \chi_\ell(k,r)} 
\hbox{\qquad with \qquad}
\chi_\ell(k,r) = \frac{rk}{i GM}
\sqrt{\left(\ell+\frac{1}{2}\right)^2\left(1-\frac{2GM}{r}\right)
\left(\frac{i}{k}\right)^2 + r^2}
\end{equation}
so that Eq.\ (\ref{eqn:kaxis2}) becomes
\begin{eqnarray}
\langle\Phi(r)^2\rangle_B
&=& \frac{1}{32\pi^2 r^2} \int_{-\infty}^\infty
\frac{dk}{\sqrt{k^2}} \left[\sum_{\ell=0}^\infty (2\ell + 1) 
\left(\rho_\ell(k,r) - \rho_\ell^{\rm WKB}(k,r)\right)
\right. \cr && \qquad \left.
+\sum_{\ell=0}^\infty \left((2\ell + 1) \rho_\ell^{\rm WKB}(k,r)
- \frac{4kr}{i\sqrt{1-\frac{2 GM}{r}}}\right)
- \frac{4 k^2 r^2}{1-\frac{2 G M}{r}}\right] 
-\frac{(GM)^2}{48 \pi^2 r^4 (1-\frac{2 G M}{r})} \,.
\label{eqn:kaxis3}
\end{eqnarray}
As shown in Ref.\ \cite{Candelas2}, by turning the sum over $\ell$
into a contour integral, we obtain
\begin{equation}
u(k,r)= \sum_{\ell=0}^\infty 
\left((2\ell + 1) \rho^{\rm WKB}_\ell(k,r) - \frac{4kr}
{i\sqrt{1-\frac{2 GM}{r}}}
\right)
- \frac{4 k^2 r^2}{1-\frac{2 G M}{r}} =
\frac{8kr}{i\sqrt{1-\frac{2GM}{r}}}
\int_0^{\Lambda(k,r)} \frac{\lambda\,  d\lambda}{(e^{2\pi\lambda} + 1)
\sqrt{\Lambda(k,r)^2 - \lambda^2}}
\end{equation}
with
\begin{equation}
\Lambda(k,r) = \frac{k}{i} \frac{r}{\sqrt{1-\frac{2 GM}{r}}} \,,
\end{equation}
and so we have
\begin{equation}
\langle\Phi(r)^2\rangle_B
= \frac{1}{32\pi^2 r^2} \int_{-\infty}^\infty
\frac{dk}{\sqrt{k^2}} \left[u(k,r) + \sum_{\ell=0}^\infty (2\ell + 1) 
\left(\rho_\ell(k,r) - \rho_\ell^{\rm WKB}(k,r)\right)\right]
-\frac{(GM)^2}{48 \pi^2 r^4 (1-\frac{2 G M}{r})} \,.
\label{eqn:kaxis4}
\end{equation}
Closing the contour in the upper-half plane, we can then convert this
expression into an integral along the branch cut on the imaginary axis
$k=i\kappa$, 
\begin{equation}
\langle\Phi(r)^2\rangle_B
= \frac{1}{32\pi^2 r^2} \int_{0}^\infty
\frac{d\kappa}{\kappa} \left[u(i\kappa,r) + \sum_{\ell=0}^\infty (2\ell + 1) 
\left(\rho_\ell(i\kappa,r) - \rho_\ell^{\rm WKB}(i\kappa,r)\right)\right]
-\frac{(GM)^2}{48 \pi^2 r^4 (1-\frac{2 G M}{r})} \,.
\label{eqn:kaxis5}
\end{equation}
Although this form can in principle be used directly for numerical
calculations, an additional subtraction introduced in Ref.\
\cite{Candelas2} significantly improves its convergence.  This change
consists of subtracting the leading contribution to the second-order
WKB approximation along with the first order WKB subtraction, and
again adding this contribution back as a contour integral, which can
be also expressed in a numerically tractable form.  Doing so, we can
use the identity \cite{Candelas2}\footnote{Note that the 
second term on the right-hand-side of Eq.\ (3.7) of Ref.\
\cite{Candelas2} should enter with a minus sign; with this change, the
first term on the right-hand side of this equation can be rewritten as
the same integral as the second term but with $\kappa=0$, and the
subtraction can then be carried out under the integral sign.  Doing so
yields the result given here, and shows explicitly that the combined
expression does not give rise to any singularities at $\kappa=0$.}
\begin{equation}
\sum_{\ell=0}^\infty (2\ell + 1) 
\frac{2r^2\kappa}{8GM \chi_\ell(i\kappa,r)^3}
=-\frac{(GM)^2}{r^2 \left(1-\frac{2GM}{r}\right)}
{\, \rm Re \,}\int_0^\pi
\frac{\tanh \left[\frac{(e^{i\theta}-1) \pi \kappa r}
{\sqrt{1-\frac{2GM}{r}}}\right]\sin \frac{\theta}{2}}
{\left(2-e^{i\theta}\right)^{3/2}} \, d\theta
\end{equation}
to obtain
\begin{equation}
\langle\Phi(r)^2\rangle_B
= \frac{1}{32\pi^2 r^2} \int_{0}^\infty
\frac{d\kappa}{\kappa} \left[\tilde u(i\kappa,r) +
\sum_{\ell=0}^\infty (2\ell + 1) 
\left(\rho_\ell(i\kappa,r) - 
\tilde \rho_\ell^{\rm WKB}(i\kappa,r)\right)\right]
-\frac{(GM)^2}{48 \pi^2 r^4 (1-\frac{2 G M}{r})} \,,
\label{eqn:kaxis6}
\end{equation}
with
\begin{equation}
\tilde \rho_\ell^{\rm WKB}(i\kappa,r) = \frac{2r^2\kappa}
{GM\chi_\ell(i\kappa,r)} \left(1+\frac{1}{8\chi_\ell(i\kappa,r)^2}\right)
\end{equation}
and
\begin{eqnarray}
\tilde u(i\kappa,r) &=& \sum_{\ell=0}^\infty 
\left((2\ell + 1) \tilde \rho^{\rm WKB}_\ell(i\kappa,r) -
\frac{4\kappa r}{\sqrt{1-\frac{2 GM}{r}}}\right)
+ \frac{4 \kappa^2 r^2}{1-\frac{2 G M}{r}}  \cr
&=& 
\frac{4 r^2 \kappa^2}{1-\frac{2GM}{r}}
\int_0^{\pi} \frac{\sin \frac{\theta}{2} \, d\theta}
{\exp\left(\frac{2 \pi \kappa r}{\sqrt{1-\frac{2GM}{r}}}
\sin \frac{\theta}{2}\right) + 1}
-\frac{(GM)^2}{r^2 \left(1-\frac{2GM}{r}\right)}
{\, \rm Re \,}\int_0^\pi
\frac{\tanh \left[\frac{(e^{i\theta}-1) \pi \kappa r}
{\sqrt{1-\frac{2GM}{r}}}\right]\sin \frac{\theta}{2}}
{\left(2-e^{i\theta}\right)^{3/2}} \, d\theta \,,
\end{eqnarray}
where we have also substituted $\displaystyle \lambda = 
\frac{\kappa r}{\sqrt{1-\frac{2 GM}{r}}} \sin \frac{\theta}{2}$
in the first term to bring the integrals into a similar form.  As
described in Ref.\ \cite{Candelas2}, these integrals are then
straightforward to compute numerically.

\section{Computational Scattering Theory Formalism}

In the standard scattering basis of Eq.\ (\ref{eqn:regs}), we can
write the spectral density in terms of regular and outgoing solutions as
\begin{equation}
\rho_\ell(k,r_*) = 
f_{+,\ell}(k,r_*) \overleftarrow{\phi_\ell}(k, r_*) + 
f_{-,\ell}(-k,r_*) \overrightarrow{\phi_\ell}(k, r_*) \,.
\end{equation}
However, especially for complex $k$, this form can lead to problems
in numerical calculation, because it requires integrating one of the
outgoing wave solutions through the peak of the Regge-Wheeler
potential, which for large $\ell$ becomes strongly repulsive.
Although this feature presents a challenge for any numerical
computation, the numerical behavior can be improved by switching from
a basis consisting of left- and right-moving wavefunctions to a basis
of symmetric and antisymmetric wavefunctions.  In this approach, the
outgoing wave solution and the regular solution are each given by a
single $2\times 2$ matrix-valued function that is defined only for
$r_* \ge z_{*,\ell}$.  The four entries of these matrices then represent the
symmetric and antisymmetric components of the wavefunctions in the
symmetric and antisymmetric scattering channels.  The asymmetric potential
is then decomposed into a symmetric component, which acts within the
symmetric and antisymmetric channels, and an antisymmetric component,
which mixes the two channels.  Analogously to a spherical problem,
where the mixing would be given in terms of Wigner 3-$j$ symbols, this
mixing is governed by the overlap integrals of three normalized
wavefunctions, which in this case are all equal to zero or $\pm 1$.

The key advantage of this basis is that the variable phase calculation
can then be reformulated in terms of a combination of the outgoing wave
solution, which is integrated in from infinity, and the regular
solution, which is integrated out from the origin
\cite{Graham:2002xq,Forrow:2012sp,Graham:2014uca}.  The resulting
improvement in the numerical behavior makes the calculations of the
previous section tractable, although in practice it remains helpful to
use quadruple precision arithmetic.  The detailed algorithm is given
in the Appendix.  Using this approach, in Fig.\ \ref{fig:delta} we graph 
\begin{eqnarray}
\Delta(r) &=& 
(2GM)^2 \left(\langle\Phi(r)^2\rangle_B
+\frac{(GM)^2}{48 \pi^2 r^4 (1-\frac{2 G M}{r})}\right) \cr
&=& \frac{(2GM)^2}{32\pi^2 r^2} \int_{0}^\infty
\frac{d\kappa}{\kappa} \left[\tilde u(i\kappa,r) +
\sum_{\ell=0}^\infty (2\ell + 1) 
\left(\rho_\ell(i\kappa,r) - 
\tilde \rho_\ell^{\rm WKB}(i\kappa,r)\right)\right] \,,
\label{eqn:kaxis7}
\end{eqnarray}
which gives a dimensionless representation of the right-hand
side of Eq.\ (\ref{eqn:kaxis6}) without the finite renormalization;
this term remains finite at the event horizon $r=2GM$.

\begin{figure}
\includegraphics[width=0.45\linewidth]{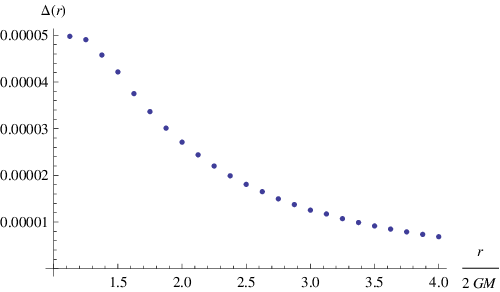}
\caption{Plot of $ \Delta(r) = 
(2GM)^2 \left(\langle\Phi(r)^2\rangle_B +\frac{(GM)^2}{48 \pi^2 r^4
(1-\frac{2 G M}{r})}\right)$ as a function of $r$, in units where
$2GM = 1$.}
\label{fig:delta}
\end{figure}

\begin{figure}
\includegraphics[width=0.45\linewidth]{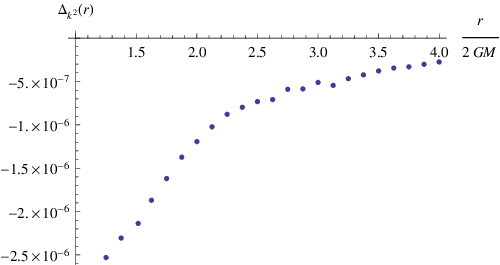}
\caption{Plot of $ \Delta_{k^2}(r) = 
(2GM)^4 \left(\langle \dot \Phi(r)^2\rangle_B +\frac{(GM)^3 (4r - 7GM)}
{480 \pi^2 r^8 (1-\frac{2 G M}{r})}\right)$ as a function of $r$, in
units where $2GM = 1$.}
\label{fig:deltak2}
\end{figure}

By applying $-\frac{d^2}{d\epsilon^2}$ before taking
$\epsilon \to 0$ in Eq.\ (\ref{eqn:pointsplit}), we can similarly
write the expectation value of $\langle\dot \Phi(r)^2\rangle_B$, where
dot denotes the derivative with respect to $t$, as
\begin{equation}
\langle\dot \Phi(r)^2\rangle_B = \frac{1}{16\pi^2 r^2} \int_0^\infty
k dk \left[\sum_{\ell=0}^\infty (2\ell + 1)\left(
|\overrightarrow{\phi}_\ell(k,r)|^2 + |\overleftarrow{\phi}_\ell(k,r)|^2
\right) - \frac{4 k^2 r^2}{1-\frac{2 G M}{r}}\right]
+\frac{(GM)^3 (4r -7GM)}{480 \pi^2 r^8 (1-\frac{2 G M}{r})}
\label{eqn:phidotsq}\,,
\end{equation}
which we can then calculate via the same approach, yielding
\begin{equation}
\langle\dot \Phi(r)^2\rangle_B
= -\frac{1}{32\pi^2 r^2} \int_{0}^\infty
\kappa d\kappa \left[\tilde u(i\kappa,r) +
\sum_{\ell=0}^\infty (2\ell + 1) 
\left(\rho_\ell(i\kappa,r) - 
\tilde \rho_\ell^{\rm WKB}(i\kappa,r)\right)\right]
+\frac{(GM)^3 (4r -7GM)}{480 \pi^2 r^8 (1-\frac{2 G M}{r})} \,.
\label{eqn:kaxisphidotsq}
\end{equation}

Again in this case
it is helpful to define a quantity proportional to the expectation
value without the finite renormalization,
\begin{eqnarray}
\Delta_{k^2}(r) = (2GM)^4 \left(\langle\dot\Phi(r)^2\rangle_B
-\frac{(GM)^3(4r - 7GM)}{480 \pi^2 r^8 (1-\frac{2 G M}{r})}\right) \,,
\end{eqnarray}
which is shown in Fig. \ref{fig:deltak2}.  Other contributions
to the stress-energy tensor \cite{PhysRevLett.53.403}
are given by similar calculations, but those cases require additional
modifications of the WKB subtraction procedure used above.

\section{Discussion}

We have shown how to use a multichannel variable phase method to
compute scattering data for Regge-Wheeler potentials arising from
quantum fluctuations in Schwarzschild spacetime.  In each
channel, this approach uses a basis of symmetric and antisymmetric
waves, which are coupled by the asymmetric potential.  These
techniques are applicable to a broad range of problems involving wave
scattering in an asymptotically flat curved spacetime background,
including cases where the frequency becomes complex, such as
quasi-normal modes or thermal field theory.  This approach can also
potentially be applied to other geometries that still allow
for a partial wave decomposition of the fluctuation spectrum, such as
cosmic string spacetimes \cite{PhysRevD.42.2669}.  Extensions of this
calculation to the stress-energy tensor could also offer the
opportunity to explore energy condition violation in the Schwarzschild
background \cite{Viss96a,Viss96b,Viss96c,Viss97a}; 
pointwise conditions on $\langle T_{\mu\nu}\rangle$ that are strong
enough to rule out exotic phenomena such as closed timelike curves,
traversable wormholes, and superluminal communication 
\cite{PhysRevLett.14.57,PhysRevLett.15.689,1979GReGr..10..985T,
PhysRevD.33.3526} are typically violated by quantum effects, but
motivated by quantum inequalities \cite{PhysRevD.43.3972,Ford:1994bj}
and information-theoretic arguments, weaker averaged
conditions may remain viable in flat 
\cite{Graham:2005cq,Fewster:2006uf,Kelly:2014mra,Bousso:2015wca,
Hartman:2016lgu} and curved
\cite{Graham:2007va,Wall:2009wi,Kontou:2012ve,Kontou:2015yha}
spacetime.

Because scattering theory methods are restricted to asymptotically flat
spacetimes, they apply to cases in which it is possible to consider
global quantities as well.  They are given in terms of the global
density of states, which in turn is related to the scattering phase
shift in each channel.  As shown in the Appendix, computing this
quantity requires greater care than in standard scattering theory
because one cannot easily separate the angular momentum barrier from
the potential; as a result, the potential falls off only as the
inverse square of distance at large positive $r_*$, and the additional
phase shift due to the angular momentum barrier, represented by the
phase difference of $\ell \pi/2$ between the spherical Hankel function
and the ordinary exponential at large distances, shows up as a
modification of Levinson's theorem for the phase shift at $k=0$.

\section{Acknowledgments}

It is a pleasure to thank E.\ N.\ Blose and N.\ Sadeh for assistance
in an earlier stage of this project, and R.\ L.\ Jaffe, M.\ Kardar,
and K.\ Olum for discussions.  N.\ G.\ was supported in part by the
National Science Foundation (NSF) through grants PHY-1520293 and PHY-1820700.

\appendix

\section{Symmetric/Antisymmetric wave basis}

In this Appendix we describe the technical details of the scattering
calculation using the basis of symmetric and antisymmetric
wavefunctions around $z_{*,\ell}$.  In this approach, all of the 
scattering functions become $2\times 2$ matrices, defined only for
$r_* \ge z_{*,\ell}$, with the actual wavefunction given by one-half the sum
of the symmetric and antisymmetric components for $r_* > z_{*,\ell}$ and
one-half the difference for $r_* < z_{*,\ell}$.  In this basis, the
asymmetric Regge-Wheeler potential contains off-diagonal terms mixing
these two channels, analogously to the case of asymmetric objects in a
spherical basis \cite{Forrow:2012sp} (the analogs of the 3-$j$
symbols are zero or $\pm 1$).  Again using the variable phase
method, we now obtain a single matrix-valued outgoing wave solution
and, by applying boundary conditions at $z_{*,\ell}$, the corresponding 
matrix-valued regular solution.  A key advantage of
this approach is that we can take advantage of the inward/outward
parameterization developed in Ref.\ \cite{Graham:2002xq}.

We take as the arbitrary ``origin'' the value of $r_*$ corresponding
to the leading estimate for the maximum of the Regge-Wheeler potential,
$z_{*,\ell} = r_*(r_{\ell}^{\rm peak})$, where \cite{Boonserm:2008zg} 
\begin{equation}
r_{\ell}^{\rm peak} = 
3 GM \left(1-\frac{1}{9 \ell(\ell+1)} \right) 
\hbox{\quad for $\ell \neq 0$, and \quad}
r_{0}^{\rm peak} = \frac{8GM}{3}
\hbox{\quad for $\ell = 0$.}
\end{equation}
We work with matrix wavefunctions $\hat \psi_\ell(k,r_*)$
that are defined only for $r_* \ge z_{*,\ell}$; to obtain the actual
scalar wavefunction, we project out the value on the left or right of
the ``origin'' by
\begin{equation}
\psi_\ell(k, r_*) = 
\left\{ \begin{array}{ll} 
{\, \rm tr \,} \left[\hat P_- \hat \psi_\ell(k,2z_{*,\ell} - r_*)\right] 
& r_* < z_{*,\ell} \\
{\, \rm tr \,} \left[\hat P_+ \hat \psi_\ell(k,r_*)\right] & 
r_* \ge z_{*,\ell} \\
\end{array}\right. \,,
\label{eqn:proj}
\end{equation}
with $\hat P_\pm = \frac{1}{2}\begin{pmatrix}
1 & \pm 1 \cr \pm 1 & 1 \end{pmatrix}$.

Following \cite{Graham:2002xq,Forrow:2012sp,Graham:2014uca}, we
parameterize the matrix outgoing wave solution to Eq.\ (\ref{eqn:KG2}) as
\begin{equation}
\hat \psi_\ell^{\rm out}(k,r_*) = e^{ik r_*} \hat g_\ell(k,r_*) \,,
\end{equation}
where we have factored out the free wavefunction $e^{ik r_*}$.
Then $\hat g_\ell(k,r_*)$ obeys the differential equation
\begin{equation}
-\hat g_\ell''(k,r_*) - 2ik \hat g_\ell'(k,r_*) +\hat v_\ell(r_*) \hat
g_\ell(k,r_*) = 0 \,,
\end{equation}
where prime denotes derivative with respect to $r_*$ and the elements
of the potential matrix are given by
\begin{equation}
\hat v_\ell(r_*)_{bc} = \sum_{a=1}^{2} 
\left(\frac{V_\ell(r_*) + (-1)^a  V_\ell(2z_{*,\ell}-r_*)}{2}\right)
\left(\frac{(-1)^{a+b+c} + 1}{2} \right) \,.
\end{equation}
Here $a=1$ is the antisymmetric channel and $a=2$ is the symmetric
channel.  Imposing outgoing wave boundary conditions for $r_* \to
\infty$, we formally would have $\hat g_\ell(k,r_* \to \infty) = 
\hat 1$ and  $\hat g_\ell'(k,r_* \to \infty) = \hat 0$.  However,
because of the slow falloff of the potential, for the large but finite
$r_*$ that arise in numerical calculations, it is advantageous to use a
better estimate of the outgoing wavefunctions for large $r_*$.  On the
right the wavefunction approaches the spherical Hankel function
solution $i^{\ell+1} k r_* h^{(1)}_\ell(k r_*)$, while on the left the
wavefunction approaches the exponential solution used above.  As a
result, in the matrix formalism a better estimate at $r_* = R_*$ 
for large positive $R_*$ is given by
\begin{equation}
\hat g_\ell(k,R_*) = 
\frac{1}{2}
\left[\frac{i^{\ell+1}k R_* h^{(1)}_\ell(k R_*)}{ e^{ik R_*}} 
\begin{pmatrix}1 & 1 \cr 1 & 1\end{pmatrix}  + 
\begin{pmatrix}1 & -1 \cr -1 & 1\end{pmatrix}\right]\,,
\end{equation}
and correspondingly, after some simplification,
\begin{equation}
\hat g_\ell'(k,R_*) = \frac{i^{\ell+1}k \left(
(1 - i k R_* + \ell) h^{(1)}_\ell(k R_*) - k R_* h^{(1)}_{\ell+1}(k R_*)
\right)}{2 e^{ik R_*}} \begin{pmatrix}1 & 1 \cr 1 & 1\end{pmatrix} \,.
\end{equation}

Again following Refs.\ \cite{Graham:2002xq,Forrow:2012sp,Graham:2014uca}, 
we can parameterize the regular solution as
\begin{equation}
\hat \psi_\ell^{\rm reg}(k,r_*) = e^{-ik(r_*-z_{*,\ell})} 
\hat h_\ell(k,r_*) \,,
\end{equation}
where $\hat h_\ell(k,r_*)$ obeys the differential equation (note the
reversed order of the matrix multiplication in the potential term)
\begin{equation}
-\hat h_\ell''(k,r_*) + 2ik \hat h_\ell'(k,r_*) +
\hat h_\ell(k,r_*) \hat v_\ell(r_*) = 0 \,.
\end{equation}
The regular solution obeys boundary conditions at the ``origin''
$z_{*,\ell}$ such that the function and first derivative match the
solution in the absence of the potential,
\begin{equation}
\hat h_\ell^{\rm free}(k,r_*) = \frac{e^{ik(r_*-z_{*,\ell})}}{k} 
\begin{pmatrix}
\sin k(r_*-z_{*,\ell}) & 0 \cr 0 & \frac{1}{i} \cos k(r_*-z_{*,\ell})
\end{pmatrix} \,.
\end{equation}

In terms of these functions, the matrix-valued local spectral density
in the symmetric/antisymmetric matrix basis is given by
\begin{equation}
\hat \rho_\ell(k,r_*) = 4 i k \hat F_\ell(k)^{-1}
\hat h_\ell(k, r_*) \hat g_\ell(k, r_*) \,,
\end{equation}
where the matrix-valued Jost function $\hat F_\ell(k)$
is given in terms of the Wronskian of the
regular and outgoing solutions as
\begin{equation}
\hat F_\ell(k) = 2ik \hat h_\ell(k, x_*) \hat g_\ell(k, x_*) + 
\hat h_\ell(k, x_*) \hat g_\ell'(k, x_*) 
- \hat h'(k, x_*) \hat g_\ell(k, x_*)\,,
\end{equation}
which can be evaluated at any fitting point $x_*$; we choose 
$x_* = r_*$ so that no new calculation is necessary.  
The scalar spectral density of Eq.\ (\ref{eqn:specden})
is then obtained using the same projection procedure as that described
in Eq.\ (\ref{eqn:proj}) above for the scattering
wavefunctions.  We note that on the imaginary $k$-axis, the Jost
function matrix can become ill-conditioned for large $\ell$, so we
compute the matrix inverse using a singular value decomposition.

Although it is not needed in the calculation of local densities, we
can also use the Jost function to obtain the $S$-matrix as
\begin{equation}
\hat S_\ell(k)  = \hat F_\ell(k)^{-1} \hat M \hat F_\ell(-k) \hat M
\end{equation}
where $\hat M$ is the constant matrix $\hat
M=\begin{pmatrix} -1 & 0 \\ 0 & 1 \end{pmatrix}$.  We can then obtain
the total phase shift as
\begin{equation}
\delta_\ell(k) = \frac{1}{2i} \log \det \hat S_\ell(k)
\end{equation}
which is in turn related to the global density of states by
\begin{equation}
\rho_\ell(k) = \frac{1}{\pi} \frac{d\delta_\ell}{dk} \,.
\end{equation}
In order to resolve ambiguities in the branch cut of the logarithm, we
define $\beta_\ell(k,r_*) = \log \det \hat g_\ell(k,r_*)$, which we
can calculate by solving
\begin{equation}
\beta_\ell'(k,r_*) = \frac{1}{2i} {\, \rm tr \,} 
\hat g_\ell'(k,r_*) \hat g_\ell(k,r_*)^{-1}
\end{equation}
subject to the boundary condition $\beta_\ell(k, R_*) = 0$.  Then by
choosing the branch of the logarithm such that
$|\delta_\ell(k) - \beta_\ell(k, z_{*,\ell})| < \pi$, we obtain a continuous
phase shift that goes to zero for $k\to \infty$.  We note that at
$k=0$, the phase shift goes to $-\pi(\ell+1)/2
$.  This result differs from the behavior of a standard
potential in one dimension, which would approach $-\pi/2$ for a
repulsive potential, because of the leading $\ell(\ell+1)/r_*^2$
behavior of the potential at large positive $r_*$; the difference of
$\ell \pi/2$ corresponds to the phase shift of the spherical Hankel
function compared to the ordinary exponential.

\bibliographystyle{apsrev}
\bibliography{gr}

\end{document}